\documentclass[12pt]{article}
\usepackage{amssymb}
\usepackage{epsfig}
\usepackage{float}
\usepackage{graphicx}

\newcommand{\be}{\begin{eqnarray}}
 \newcommand{\ee}{\end{eqnarray}}
 \newcommand{\nee}{\nonumber\end{eqnarray}}
 \newcommand{\nn}{\nonumber\\}

\def\m             {\mu}
 \def\n              {\nu}
\def\t          {\tau}

\begin{document}
\begin{center}
{\bf On the polarization of the final nucleon in \\ NC elastic
$\nu_{\mu}(\bar\nu_{\mu})-N$
 scattering }
\end{center}
\vspace{.7cm}

\begin{center}
{ \bf S. M. Bilenky} \vspace{.1cm}

 {\em  Joint Institute for Nuclear
Research, Dubna, R-141980, Russia;\\}

{\em TRIUMF 4004 Wesbrook Mall, Vancouver BC, V6T 2A3 Canada\\}

\end{center}

\begin{center}
{\bf  E. Christova} \vspace{.1cm}

{\em  Institute for  Nuclear Research and Nuclear Energy of BAS,\\ Sofia 1784, Bulgaria\\}
\end{center}
\vspace{.1cm}

\begin{abstract}
New short baseline neutrino experiments open new
possibilities of high precision study of different neutrino
processes. We present here
 results of the calculation of
the  polarization of final nucleon in elastic NC $\nu_{\mu}$ ($\bar\nu_{\mu}$ )-nucleon scattering.
 In a numerical analysis the sensitivity to the different choices of the
 axial and axial strange form factors is examined. Measurements of the polarization of the final  proton in elastic
$e-p$ scattering drastically changed our knowledge about  the
electromagnetic form factors of the proton. From  measurement of the
nucleon polarization  in the NC elastic scattering a new additional
information about the axial  $G_{A}(Q^{2})$ and the strange axial
 $G^{s}_{A}(Q^{2})$ form factors  of the nucleon could be inferred.
\end{abstract}

\section{Introduction}
The study of the Neutral Current (NC) elastic scattering of neutrino and antineutrino on the nucleon
\begin{equation}\label{NC}
\nu_{\mu}+N\to \nu_{\mu}+N
\end{equation}
\begin{equation}\label{1NC}
\bar\nu_{\mu}+N\to \bar\nu_{\mu}+N
\end{equation}
  is an important source of information on weak form factors of the nucleon.
  The effective Standard Model (SM) Lagrangian of these processes has the form
\begin{equation}\label{NC1}
    \mathcal{L}_{I}=-\frac{G_{F}}{\sqrt{2}}\bar\nu_{\mu}\gamma^{\alpha}
    (1-\gamma_{5})\nu_{\mu}~j_{\alpha}^{NC}.
\end{equation}
Here
\begin{equation}\label{NC2}
 j_{\alpha}^{NC}=2j_{\alpha}^{3} -2\sin^{2}\theta_{W} j_{\alpha}^{EM}
\end{equation}
is the  NC of quarks. In Eq.(\ref{NC2}) $\theta_{W}$ is the weak angle,
$j_{\alpha}^{EM}$ is the electromagnetic current of quarks and
\begin{equation}\label{NC3}
 j_{\alpha}^{3}=\sum^{3}_{a=1}\bar\psi_{aL}\gamma_{\alpha}\frac{1}{2}\tau_{3}   \psi_{aL}
\end{equation}
is the third component of the isovector current ($\psi_{aL}~ (a=1,2,3)$ is the left-handed quark doublet).

For the matrix element of the process (\ref{NC}) we have
\begin{equation}\label{Matelem}
 \langle f|(S-1)|i\rangle =-i\frac{G_{F}}{\sqrt{2}}N_{k'} N_{k}
 \bar u(k')\gamma^{\alpha} (1-\gamma_{5}) u(k)~ _{p}\langle p'|J^{NC}_{\alpha} |p\rangle_{p} (2\pi)^{4}\delta(p'-p-q).
\end{equation}
Here $k$ ($p$) and $k'$ ($p'$) are momenta of the initial
and final neutrinos (nucleons), $q=k-k'=p'-p$ is the momentum transfer
$J^{NC}_{\alpha}$ is the hadronic Neutral Current in the Heisenberg
representation and $N_{k}=(\frac{1}{(2\pi)^{3}2k^{0}})^{1/2}$ is the standard normalization factor.

We will take into account the light $u,d,s$ quarks. In this case, for the SM neutral
current we have:
\begin{equation}\label{NC4}
J^{NC}_{\alpha}=(V^{3}_{\alpha}-A^{3}_{\alpha}) -\frac{1}{2}(V^{s}_{\alpha}-A^{s}_{\alpha})-2\sin^{2}\theta_{W} J_{\alpha}^{EM}.
\end{equation}
Here $V^{3}_{\alpha}$ and $A^{3}_{\alpha}$ are third components of isovector currents $V^{i}_{\alpha}$ and $A^{i}_{\alpha}$,
$V^{s}_{\alpha}$ and $A^{s}_{\alpha}$ are strange vector and axial currents. The  electromagnetic current is connected with $V^{3}_{\alpha}$ by the relation
\begin{equation}\label{NC5}
 J_{\alpha}^{EM}=V^{3}_{\alpha}+ V^{0}_{\alpha},
\end{equation}
where $V^{0}_{\alpha}$ is the isoscalar current.

Using charge symmetry from (\ref{NC5}) for the one nucleon matrix elements we find
\begin{equation}\label{NC6}
  _{p,n}\langle p'|V^{3}_{\alpha} |p\rangle_{p,n}=\pm\frac{1}{2} \left [
_{p}\langle p'|J^{EM}_{\alpha} |p\rangle_{p}-
_{n}\langle p'|J^{EM}_{\alpha} |p\rangle_{n}\right].
\end{equation}
The isovector $A^{i}_{\alpha}$ satisfies the relation
\begin{equation}\label{NC7}
[A^{i}_{\alpha}, T_{k}]=ie_{ikl}A^{k}_{\alpha},
\end{equation}
where $T_{k}$ is the operator of the total isotopic spin.
From this relation we find
\begin{equation}\label{NC8}
 _{p,n}\langle p'|A^{3}_{\alpha} |p\rangle_{p,n}=\pm\frac{1}{2}
~_{p}\langle p'|A^{1+i2}_{\alpha} |p\rangle_{n}.
\end{equation}

The one-nucleon matrix elements of the electromagnetic current are given by the relations
\begin{equation}\label{EM1}
_{p,n}\langle p'|J^{EM}_{\alpha} |p\rangle_{p,n}=
N_{p'} N_{p}\bar u(p')[\gamma^{\alpha}F^{p.n}_{1}(Q^{2})+\frac{i}{2M}  \sigma_{\alpha\beta}q^{\beta}F^{p.n}_{2}(Q^{2})]u(p),
\end{equation}
where $F^{p,n}_{1}(Q^{2})$ and $F^{p,n}_{2}(Q^{2})$ are the Dirac and Pauli form factors of the proton (neutron) and $Q^{2}=-q^{2}$.

For the matrix element of the CC axial current $A^{1+i2}_{\alpha}$ we have:
\begin{equation}\label{CC1}
_{p}\langle p'|A^{1+i2}_{\alpha} |p\rangle_{n}=
N_{p'} N_{p}\bar u(p')[\gamma^{\alpha}\gamma_{5} G_{A}(Q^{2})+\frac{1}{2M}\gamma_{5} q^{\alpha}G_{P}(Q^{2})]u(p),
\end{equation}
where $G_{A}(Q^{2})$ and $G_{P}(Q^{2})$ are the axial and pseudoscalar
form factors of the nucleon.

From time reversal invariance of  strong interactions
it follows that the matrix elements of the strange vector and axial currents have the same form as the matrix elements of $J^{EM}_{\alpha}$ and
$A^{1+i2}_{\alpha} $, respectively(see \cite{ABMaj}).

Finally,  for the one-nucleon matrix element of the neutral current we find\footnote{
It is obvious that that pseudoscalar form factors do not make contribution to
 the matrix element of processes (\ref{NC}) and (\ref{1NC})}
\begin{equation}\label{NC9}
_{p,n}\langle p'|J^{NC}_{\alpha} |p\rangle_{p,n}=
N_{p'} N_{p}\bar u(p')J_{\alpha} ^{NC(p,n)} u(p).
\end{equation}
Here
\begin{equation}\label{NNC}
 J_{\alpha} ^{NC(p,n)}=V_{\alpha} ^{NC(p,n)}-A_{\alpha} ^{NC(p,n)},   \end{equation}
where
\be
 V_{\alpha} ^{NC(p,n)}&=&\gamma_{\alpha}F^{NC(p,n)}_{1}(Q^{2})+\frac{i}{2M}  \sigma_{\alpha\beta}q^{\beta}F^{NC(p,n)}_{2}(Q^{2}),\nn
A_{\alpha} ^{NC(p,n)}&=&
\gamma_{\alpha}\gamma_{5} G^{NC(p,n)}_{A}(Q^{2}).\label{NC10}
\ee
In this relation we have
\begin{equation}\label{NC11}
F^{NC(p,n)}_{1,2}(Q^{2})=\pm\frac{1}{2}(F^{p}_{1,2}(Q^{2})-F^{n}_{1,2}(Q^{2}))
-\frac{1}{2}~F^{s}_{1,2}(Q^{2})-2\sin^{2}\theta_{W}F^{p,n}_{1,2}(Q^{2})
\end{equation}
and
\begin{equation}\label{NC12}
G^{NC(p,n)}_{A}(Q^{2})=\pm \frac{1}{2} G_{A}(Q^{2})-\frac{1}{2}~G^{s}_{A}(Q^{2}),
\end{equation}
where $F^{s}_{1,2}(Q^{2})$ and $G^{s}_{A}(Q^{2})$ are vector and axial strange form factors of the nucleon.

Thus, the matrix elements of the processes (\ref{NC}) and  (\ref{1NC})
are determined by the known electromagnetic form factors of the proton and the neutron,
the axial form factor of the nucleon $G_{A}(Q^{2})$ and the strange form factors of the nucleon.

An information on the axial form factor $G_{A}(Q^{2})$ is inferred from  study of the CC quasi-elastic (CCQE) processes:
\begin{equation}\label{CC}
\nu_{\mu}+n\to \mu^{-}+p,\quad \bar\nu_{\mu}+p\to \mu^{+}+n
\end{equation}
The axial form
factor  is usually parameterized by the dipole formula
\begin{equation}\label{dipoleGA}
G_{A}(Q^{2})=\frac{g_{A}}{(1+\frac{Q^{2}}{M^{2}_{A}})^{2}}.
\end{equation}
Here  $g_{A}\simeq 1.27$  is the axial constant,
and $M_{A}$ is a parameter ( the "axial mass").

 The values of the parameter $M_{A}$ determined from the data of different experiments under
 the assumption that impulse approximation is valid (neutrino interacts with a quasi-free nucleon in
 a nuclei and other nucleons are spectators) are quite different.

From analysis of the old bubble chamber data on  measurements of the cross section of the process $\nu_{\mu}+n\to \mu^{-} +p$
 on deuterium target and of the process  $\bar\nu_{\mu}+p\to \mu^{+} +n$
on  proton target it was found  \cite{Bodek}:
\begin{equation}\label{Bodek}
 M_{A}=1.016\pm 0.026~\mathrm{GeV}.
\end{equation}

The value of the parameter $M_{A}$ obtained from  the measurement of
 the CCQE cross section in the NOMAD experiment (carbon target) \cite{Nomad}
\begin{equation}\label{Nomad}
 M_{A}=1.05\pm 0.02 \pm 0.06~\mathrm{GeV}
\end{equation}
is in agreement with (\ref{Bodek}).

However, from  fit of the data of more recent experiments
larger average values of the parameter  $M_{A}$  were obtained.
From the data of the MINOS experiment (iron target) it was found \cite{Minos}:
\begin{equation}\label{Minos}
 M_{A}=1.26 _{-0.10}^{+0.12} {}_{-0.12}^{+0.08}~\mathrm{GeV}.
\end{equation}

 In the K2K experiment ($H_{2}O$ target) it was obtained~\cite{K2K}:
\begin{equation}\label{K2K}
 M_{A}=1.20\pm 0.12~\mathrm{GeV}.
\end{equation}

From the high-statistics MiniBooNE experiment (carbon target) it was
 inferred \cite{Miniboone}:
\begin{equation}\label{Miniboone}
 M_{A}=1.35\pm 0.17~\mathrm{GeV}.
\end{equation}

There could be many different reasons for such disagreement.
It could be a problem of systematics and normalization.
Target nuclei in the different experiments are different.
The difference of the values of  $M_{A}$ obtained from the data of different experiments
 could be due to  various nuclei effects(see \cite{Martini1, Martini2}).

The axial form factor $G_{A}(Q^{2})$ is a fundamental characteristic of the nucleon.
It is of a great theoretical interest. CCQE processes (\ref{CC}) are the dominant neutrino processes
in the GeV energy range. The modern high precision neutrino oscillation experiments
require a percentage-level knowledge of the axial form factor and cross sections of CCQE processes (\ref{CC}).
 In several dedicated neutrino experiments (T2K\cite{T2K}, MINERVA\cite{MINERVA},
 ArgoNeuT\cite{ArgoNeuT}) new measurements of CCQE cross section will be performed.

In \cite{BilChr} we proposed a  measurement of the polarization of the recoil nucleon in CCQE processes (\ref{CC}) for
a determination of the axial form factor.

A measurement of the polarization of the recoil protons in the elastic $e-p$ scattering drastically changed our understanding of
the electromagnetic form factors of the proton (see \cite{Perdrisat,Arrington}). Before these measurements were done
the results of the  analysis of the data of the numerous experiments on  measurements of the cross section of
 elastic scattering of unpolarized electrons on unpolarized protons indicated that the ratio
$R(Q^{2})$ of the electric and magnetic form factors of the proton
does not depend on $Q^{2}$ and is close to one. A measurement of the ratio of the transverse and longitudinal polarizations of the proton allows
to determine the ratio of the electric and magnetic form factors in a direct model independent way. After these
 measurements were done it was established that the ratio $R(Q^{2})$ decreased linearly with $Q^{2}$
($R\simeq 1$ at $Q^2 \simeq 1 ~\mathrm{GeV}^{2}$ and
 $R= 0.28\pm 0.09$ at $Q^{2}=5.6 ~\mathrm{GeV}^{2}$).
The measurement of the polarization significantly changed  the  theoretical models for the structure of the nucleon.

In this paper we  present the results of calculation of the polarization of the recoil nucleon in the elastic NC processes (\ref{NC}) and (\ref{1NC}).
It is natural to expect that  measurements of the polarization of the
nucleon in CCQE scattering and NC elastic scattering
could provide important information about the axial form factor of the
nucleon. From our point of view it is worthwhile  to consider a possibility of such measurements
in modern high-statistics short baseline neutrino experiments in which hundreds of thousands of neutrino events are observed.

 Here  we present the results of the calculation of the the
polarization of the final nucleon in the case of the definite
neutrino energy and free nucleon target. In order to obtain the
polarization in a realistic neutrino experiment with a spectrum of
initial neutrinos (antineutrinos) one needs to average the
expressions presented below over the spectrum. Let us notice that
 the numerator and the denominator in the expressions  (\ref{polar}) and
  (\ref{polar4}) must be averaged separately. In modern neutrino experiments
   nuclear targets such as carbon, oxygen, iron or argon are used. It was shown in many papers
    (see, for example, \cite{Martini1,Martini2})
 that nuclear effects are important and must be taken into account. We do not consider  nuclear effects here.

Investigation of the NC processes (\ref{NC}) and (\ref{1NC}) allows to obtain an
information about strange form factors of the proton (see, for example, \cite{ABMaj}).
Strange vector form factors can be inferred from experiments on the study of the P-odd asymmetry
in the elastic scattering of longitudinally polarized electrons on unpolarized proton and other targets.
From many experiments performed at different values of $Q^{2}$ it follows that strange vector form factors are small,
compatible with zero. For example, from analysis of the data of the recent HAPPEX experiment at
JLab \cite{Happex} it was found that at $Q^{2}\simeq 0.62 ~GeV^{2}$ the charge and magnetic strange form factors are:
\begin{equation}\label{strange}
    G^{s}_{E}=0.047\pm0.034,\quad  G^{s}_{M}=0.070\pm0.067.
\end{equation}
An information about  $G^{s}_{A}$ was obtained from the data of the BNL experiment  on the measurement of the cross sections of the
NC processes (\ref{NC}) and (\ref{1NC})\cite{BNL}. From analysis of the data it was found:
\be
G^{s}_{A}(0)=-0.21\pm 0.10.
\ee
There exist, however, a strong correlation between the values of the parameters $G^{s}_{A}(0)$ and $M_{A}$.
Taking into account this correlation we can conclude that $-0.25<G^{s}_{A}(0)<0$ (see, \cite{ABMaj}).

\section{Polarization of the the final nucleon in NC elastic scattering}

We will present here the result of the calculation of the
polarization of the final nucleon in NC processes (\ref{NC}) and
(\ref{1NC}). In the covariant density matrix formalism
the 4-vector of the polarization of the final nucleon produced in
(\ref{NC}) and (\ref{1NC}) is given by the expression:
\begin{equation}\label{polar}
\xi^{\tau}=\frac{\mathrm{Tr}[\gamma^{\tau}\gamma_{5}~\rho_{f}]}
{\mathrm{Tr}[\rho_{f}]}.
\end{equation}
Here the final density matrix $\rho_{f}$ is determined by the
expression
\begin{equation}\label{polar1}
 \rho_{f}=L^{\alpha\beta} \Lambda(p')J_{\alpha} ^{NC(p,n)}\Lambda(p)J_{\beta} ^{NC(p,n)}
\Lambda(p'),
\end{equation}
where
\begin{equation}\label{polar2}
L^{\alpha\beta}=\mathrm{Tr}\left[\gamma^{\alpha}(1-\gamma_{5})k\!\!\!/
\gamma^{\beta}(1-\gamma_{5})k'\!\!\!\!/\right],
\end{equation}
$J_{\beta} ^{NC(p,n)}$ is given by  (\ref{NNC}) and $\Lambda(p)=p\!\!\!/+M$. Taking into account the relation
\begin{equation}\label{polar3}
\Lambda(p')\gamma^{\tau}\gamma_{5}\Lambda(p')=2M\left(g^{\tau\sigma}-
\frac{p'^{\tau}p'^{\sigma}}{M^{2}}\right)\Lambda(p')\gamma_{\sigma}
\gamma_{5}
\end{equation}
we can rewrite the expression for the 4-vector of the polarization in the form
\begin{equation}\label{polar4}
\xi^{\tau}=\left( g^{\tau\sigma}-\frac{(p')^{\tau}(p')^{\sigma}}{M^2}\right)
\frac{L^{\alpha\beta}\mathrm{Tr}
\left[\gamma_{\sigma}\gamma_{5}\Lambda(p')J_{\alpha} ^{NC(p,n)}\Lambda(p)J_{\beta} ^{NC(p,n)}\right]}
{L^{\alpha\beta}\mathrm{Tr}\left[\Lambda(p')J_{\alpha} ^{NC(p,n)}\Lambda(p)J_{\beta} ^{NC(p,n)}\right]}
\end{equation}
After  lengthy calculations for the vector of polarization in the rest frame of
the initial nucleon we will find the following expression:
\begin{equation}\label{Polar}
\vec{\xi}=\frac{1}{J_0\,E}\left\{(\vec{k}+\vec{k'})P_{+}+
\vec{q}~\left[-\frac{E+E'}{M}~P_{+}
+(1+\frac{E-E'}{M})~\left(P_{-}-P_p\right)\right]\right\}.
\end{equation}
Here
\begin{equation}\label{Polar1}
 P_{+}= [y\,G_M^{NC}+(2-y)G^{NC}_A] \,G_E^{NC},
 \end{equation}

\begin{equation}\label{Polar2}
P_{-}-P_p=-[(2-y)\,G_{A}^{NC}+y G^{NC}_{M}]
[G_{A}^{NC}+\frac{\tau}{y}(2-y)F^{NC}_{2}]
\end{equation}
and
\begin{eqnarray}\label{Polar3}
    J_{0}^{\nu ,\,\bar\nu}
 &=&2(1-y)\left[(G^{NC}_{A})^{2}+\frac{\tau (G_{M}^{NC})^{2}+(G_{E}^{NC})^{2}}{1+\tau}\right]\nn
 &&+\frac{My}{E}\,\left[\,(G^{NC}_{A})^{2}-\frac{\tau (G_{M}^{NC})^2+(G_{E}^{NC})^2}{1+\tau}\right]\nonumber\\
 &&+y^{2}\,(G_{M}^{NC}\mp G_{A}^{NC})^2\pm 4y\,G_{M}^{NC}\,G^{NC}_{A}.
\end{eqnarray}
 The quantities  $J_{0}^{\nu ,\,\bar\nu}$ are  connected to the cross sections of the processes  (\ref{NC}) and
(\ref{1NC}) by the relations:
\begin{equation}\label{Polar4}
J_0^{\nu ,\bar\nu}=\frac{4\pi}{G_F^2}~\frac{d\sigma^{\nu ,\,\bar\nu}}{dQ^2}.
\end{equation}
In  equations (\ref{Polar})-(\ref{Polar2}) $E$ and $E'$ are the  energies of the initial and final neutrinos
in the lab. frame,
\be
y=\frac{pq}{pk},\quad  \tau=\frac{Q^{2}}{4M^{2}},\quad G_M^{NC} =F^{NC}_{1}+F^{NC}_{2},\quad  G^{NC}_{E}=F^{NC}_{1}-\tau F^{NC}_{2}.
\label{kinems}
\ee

From (\ref{Polar}) it follows that the  polarization vector lays in the scattering plane.
We expand this vector along the  two orthogonal unit vectors $\vec{e}_{\|}$ and $\vec{e}_{\bot}$ determined as follows:
\begin{equation}\label{vectors}
\vec{e}_L=\frac{\vec{p'}}{|\vec{p'}|}=\frac{\vec{q}}{|\vec{q}|},
\qquad \vec{e}_T=\vec{e}_L\times \vec{n},\qquad
\vec{n}=\frac{\vec{q}\times\vec{k}}{|\vec{q}\times\vec{k}|}.
\end{equation}
We have:
\begin{equation}\label{Polar4}
 \vec{\xi}=\xi_T\vec{e}_T+\xi_L\vec{e}_L,
\end{equation}
where $\xi_T$ and $\xi_L$ are the transverse and longitudinal polarizations. From (\ref{Polar}), (\ref{Polar1})
and (\ref{Polar2})
for the transverse polarization we obtain the  expression:
\begin{equation}\label{Polar5}
\xi^{\nu,\bar\nu}_T=-\frac{2\sin\theta_{N}}{J^{\nu,\bar\nu}_{0}}~ [\pm y\,G_M^{NC}+(2-y)G^{NC}_A] \,G_{E}^{NC},
\end{equation}
where $\theta_{N}$ is the angle between momenta of the initial neutrino and the final nucleon.

It is obvious that $\xi_T=s_T$ where
$s_T$ is the transverse polarization in the rest frame of the recoil nucleon.
For the longitudinal polarization in the rest frame of the recoil nucleon we find:
\begin{equation}\label{Polar6}
s^{\nu,\bar\nu}_L= \frac{q_0}{ |\vec
q|J^{\nu,\bar\nu}_{0}}\,\left[ \pm y\,G_{M}^{NC}+(2-y)\,G^{NC}_A\right]
 \left[(2-y)\,G_M^{NC}\pm y\left( \frac{1+\tau}{\tau}\right)\,G^{NC}_A\right].
\end{equation}

 In the case of the NC processes (\ref{NC}) and (\ref{1NC}) only the
 energy $E'_{N}$  of the final nucleon and the scattering angle
$\theta_{N}$ can be measured. In terms of these quantities we have:
\begin{equation}\label{Polar7}
Q^{2}=2M (E'_{N}-M),\quad y=\frac{(E'_{N}-M)}{E},\quad
q_0=E'_N-M,\quad \vert\vec q\vert=\sqrt{E^{'\,2}_N-M^2}.
\end{equation}
The neutrino energy $E$ is determined by the relation:
\begin{equation}\label{E}
    E=\frac{M (E'_{N}-M)}{M-E'_{N}+p'_{N}\cos\theta_{N}},\qquad p'_N =\sqrt{E^{'\,2}_N-M^2}.
\end{equation}

\section{Comments}

 Our comments are based on the following two characteristic features of the NC neutrino processes.

i) From (\ref{NC12}) it follows that the axial form factor appears
only in the combinations $G_A-G_A^s$ -- if measurements are on
protons, and $G_A+G_A^s$ -- if measurements are on neutrons.

 ii) From (\ref{NC11}) and (\ref{kinems}) we find the following expressions for the
the NC magnetic and electric form factors of the proton and the
neutron:
\begin{equation}\label{NC_p}
G_{M,E}^{NC(p)}= \frac{1}{2}(1-4\sin^{2}\theta_{W})~G_{M,E}^{p}
-\frac{1}{2}~G_{M,E}^{n}-\frac{1}{2}~G_{M,E}^{s}
\end{equation}
and
\begin{equation}\label{NC_n}
G_{M,E}^{NC(n)}=\frac{1}{2}(1-4\sin^{2}\theta_{W})~G_{M,E}^{n}
-\frac{1}{2}~G_{M,E}^{p}-\frac{1}{2}~G_{M,E}^{s},
\end{equation}
in which $\sin^2\theta_W$ enters in the combination
$(1-4\sin^2\theta_W)$. From analysis of the existing experimental
data it follows that \cite{PDG}
\begin{equation}\label{PDG}
 \sin^{2}\theta_{W}=0.23116\pm 0.00012.
\end{equation}
This  implies  that $(1-4\sin^2\theta_W)\simeq 0.075$   and,
consequently,  the NC charge form factor of the proton is
very small:
\be
G_{E\,p}^{NC}&\simeq&\,\frac{1}{2}\,\left[\,0,075\,G_E^p-\,G_E^n-G_E^s\right]\simeq
0. \label{GEp}
 \ee

$\bullet$ The   transverse polarizations of the final protons and nucleons are determined in (\ref{Polar5})
and are directly proportional to the NC charge form factors $G_E^{NC}$.

Eq. (\ref{GEp}) implies that the transverse polarization of the proton is strongly suppressed,
 which makes its measurement a very difficult task.

 Of interest could be the transverse polarization of the final neutron.
It exhibits a simple linear dependence on $G_A+G_A^s$:
\be
G_A+G_A^s&=&\frac{2}{2-y}\left[\frac{1}{2\,\sin\theta_N}\,
\frac{(J_0\,s_T)^{\n ,\,\bar\n}_{n}}{G_{E,n}^{NC}}\,\pm y\,G_{M,n}^{NC}\right]\\
&\simeq&\frac{-1}{2-y}\left[-\,\frac{1}{\sin\theta_N}\,
\frac{(J_0\,s_T)^{\n ,\,\bar\n}_{n}}{G_E^p+G_E^s}\,\pm y\,G_{M,n}^{NC}\right]
\ee
In the last line we have used:
\be
G_{E\,n}^{NC}=\frac{1}{2}\left[ 0,075 G_E^n -G_E^p -G_E^s\right]&\simeq&\,-\,\frac{1}{2}\,\left(G_E^p+G_E^s\right)
\ee
\\

$\bullet$
The longitudinal polarization is determined in (\ref{Polar6}). Note that the electric form factors
do not enter this expression and the
 longitudinal polarization is expressed only in terms of $G_M^{NC}$ and the axial form factors.
From
(\ref{NC_p}) and (\ref{NC_n}) it follows that both for   protons  and  neutrons it
is expressed entirely in terms of the best measured  magnetic form factors $G_M^{p,n}$, the small strange vector form factor $G_{M}^s$
 and the axial form factors $(G_A\pm G_A^s)$ {bf skip this prase ( that we want to determine).}

In order to measure the longitudinal polarization,
the neutrino detector must be placed in a magnetic field. 

These expressions considerably simplify forming the sum of the longitudinal polarizations of $\n$ and $\bar\n$.
Then measurements on protons and neutrons provide
 two linear equations for $(G_A\pm G_A^s)$:
\be
G_A-G_A^s&=&\frac{\sqrt{\t (1+\t )}}{[y^2(1+\t )+\t (2-y)^2]}\,
\frac{\left(J_0\,s_L\right)^{\n +\,\bar\n}_{p}}{G_{M,p}^{NC}}\label{long_p}\\
G_A+G_A^s&=&-\,\frac{\sqrt{\t (1+\t )}}{[y^2(1+\t )+\t (2-y)^2]}\,
\frac{\left(J_0\,s_L\right)^{\n +\,\bar\n}_{n}}{G_{M,n}^{NC}}\label{long_n}
\ee

$\bullet$ If  both the transverse
and longitudinal polarizations can be measured,
then we can determine their ratio (like in the case of elastic $e - p$ scattering):
\be
\left(\frac{s_L}{s_T}\right)^{\n ,\bar \n }=\frac{-\,M}{\vert\vec q\vert\sin\theta_N}\,\,
\frac{\left[\t (2-y)\,G_M^{NC}\pm y\,(1+\t )\,G_A^{NC}\right]}{G_E^{NC}}.\label{ratio}
\ee
As the transverse polarization of the proton is strongly suppresses, we consider the polarization of the final neutron only.
From (\ref{ratio}), for $G_A+G_A^s$ we obtain:
\be
G_A+G_A^s=\frac{\pm\,2}{y}\,\sqrt{\frac{\t}{1+\t}}\,\left[\frac{2\,\sin\theta_N}{G_{E,n}^{NC}}\,
\left(\frac{s_L}{s_T}\right)^{\n ,\bar \n }+\sqrt{\frac{\t}{1+\t}}\,(2-y)G_{M,n}^{NC}\right]
\ee
An advantage of  the ratio $s_L/s_T$ is that many of the systematic uncertainties cancel.

\section{Numerical analysis}

 Here we  present the results of the study of
 the sensitivity of  the transverse and longitudinal polarizations
 to the  different choices of the axial form factors $G_A$ and $G_A^s$. For comparison,
   we present also the cross sections for the same values of $G_A$ and $G_A^s$.

  We  use  the following commonly used
 parameterizations for the  form factors, summarized in \cite{Perdrisat}:
 \be
 G_D&=&\frac{1}{\left(1+\frac{Q^2}{M_V^2}\right)^2},\qquad M_V^2=0.71\nn
 G_{M,p}&=&\m_p\,G_D,\qquad  G_{M,n}=\m_n\,G_D\nn
G_{E,p}&=&(1.06-0.14\,Q^2)\,G_D\nn
G_{E,n}&=&-a\,\frac{\m_n\t}{1+b\t}\,G_D,\qquad a=1.25,\quad
b=18.3\nn G_A^s&=&\frac{g_a^s}{\left(1+\frac{Q^2}{M_A^2}\right)^2}
\nn
G_M^s&=&G_E^s=0\label{parms}
\ee
 where $\m_p=2.79$ and  $\m_n=-1.91$ are the magnetic moments of the proton and neutron.
 Our free parameters are $M_A$ and $g_a^s$. We have calculated the effect of the different axial form factors on
 the polarizations, considering the following choices of  $M_A$ and $g_a^s$:\\

1) $M_A=1.016$,\, $g_a^s=0$ -- full line\\

2) $M_A=1.016$,\, $g_a^s=-0.21$ -- dashed line\\

3) $M_A=1.35$,\, $g_a^s=0$ -- dotted line\\

4) $M_A=1.35$,\, $g_a^s=-0.21$ -- dash-dotted line\\
\\
Note that we assume the same $Q^2$-dipole form  for the axial and strange axial form factors.

We present the polarizations for  two values of the neutrino energy: $E=1$ GeV and $E=5$ GeV. The plots are given
 as functions of $Q^2$  in the interval $Q^2_{min}\leq Q^2\leq Q^2_{max}$, where
$Q^2_{min}$ is determined from $E'_N\geq M$,  and $Q^2_{max}$ --
by the condition $\cos\theta_N \leq 1$. Once we fix $E$ and $Q^2$, the scattering angle  $\theta_N$ is determined via (\ref{E}).
We have:
\be
\cos\theta_N = \sqrt{\frac{\t}{1+\t }}\, \left( 1+\frac{M}{E}\right) \leq 1
\ee
 which implies $Q^2_{max}= 4ME^2/(2E+M)$ and
 \be
 \sin\theta_N=\sqrt{\frac{Q^2}{Q^2+4M^2}\,\left(\frac{4M^2}{Q^2}-\frac{2M}{E}-\frac{M^2}{E^2}\right)}.
 \ee

 We examine separately $\n -N$ and $\bar\n -N$ elastic scattering.

$\bullet$  First we show the polarizations in $\bar\nu -N$  elastic scattering.

 \begin{figure}[htb]
 \centerline{ \epsfig{file=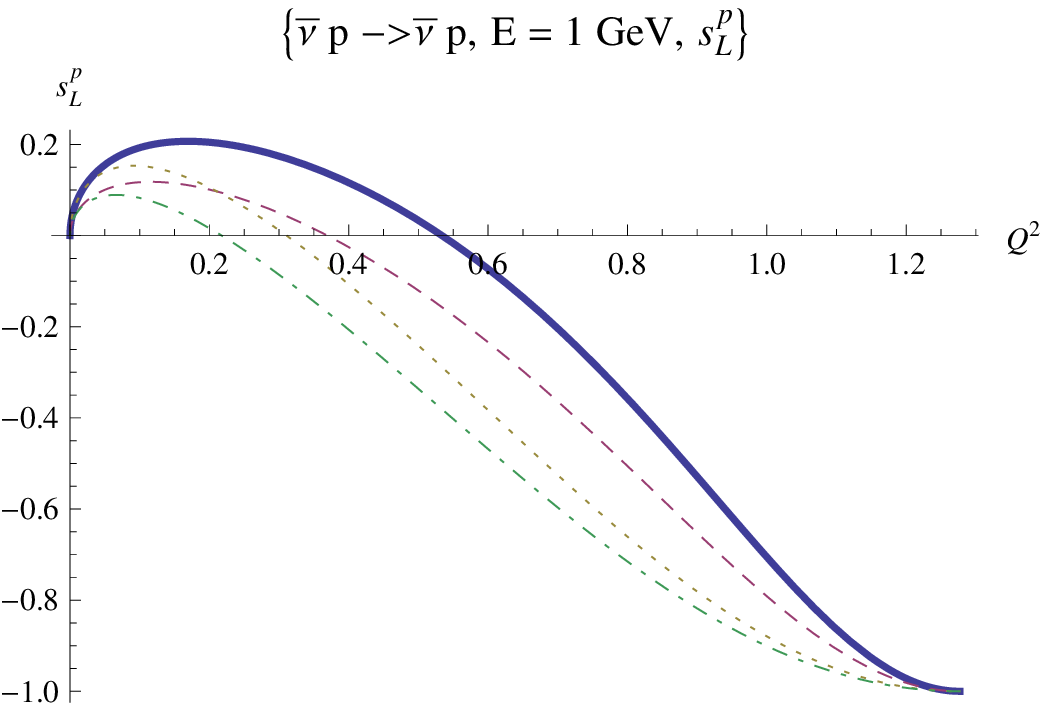,width=75mm}\hspace{0.5cm}\epsfig{file=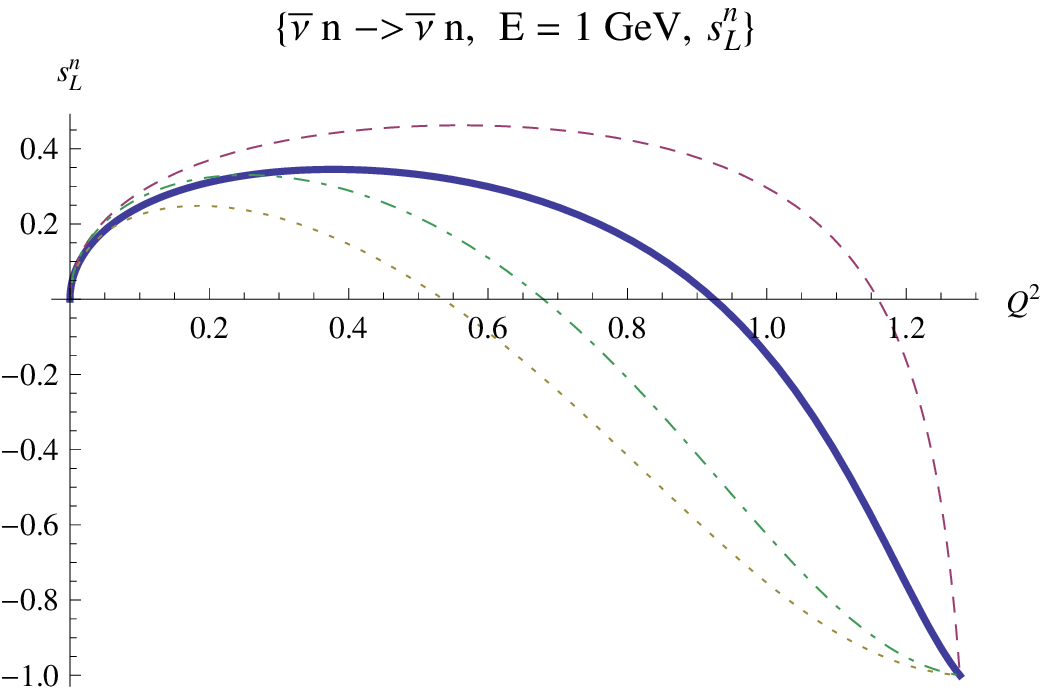,width=75mm} }
 \vspace{10mm}

\centerline{\epsfig{file=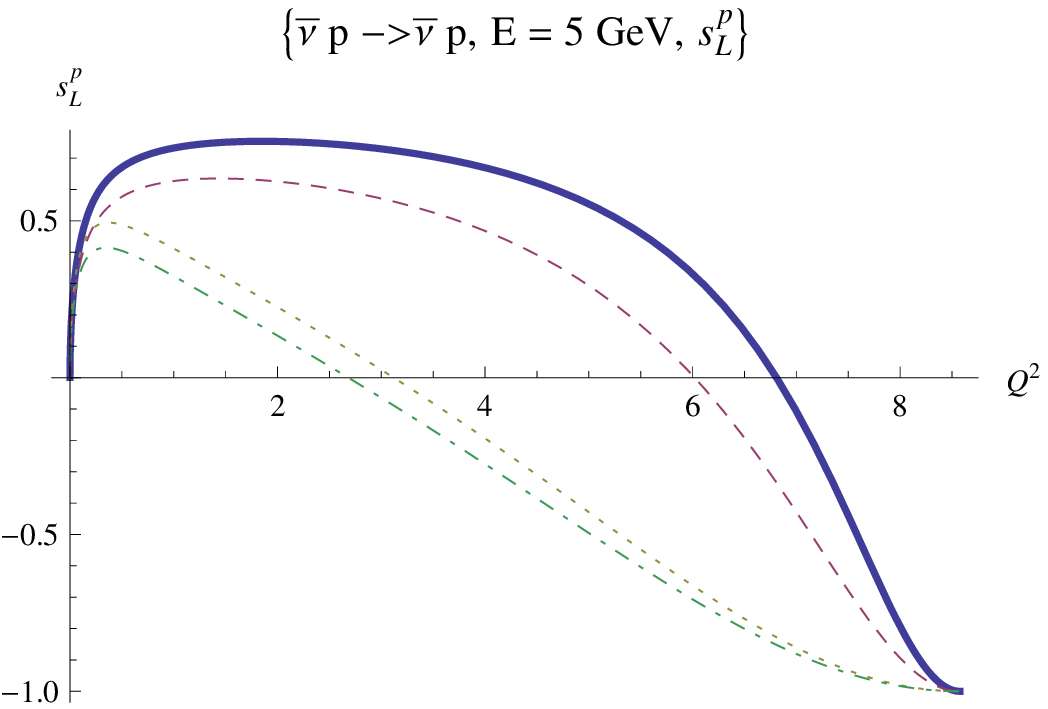,width=75mm}
\hspace{0.5cm}\epsfig{file=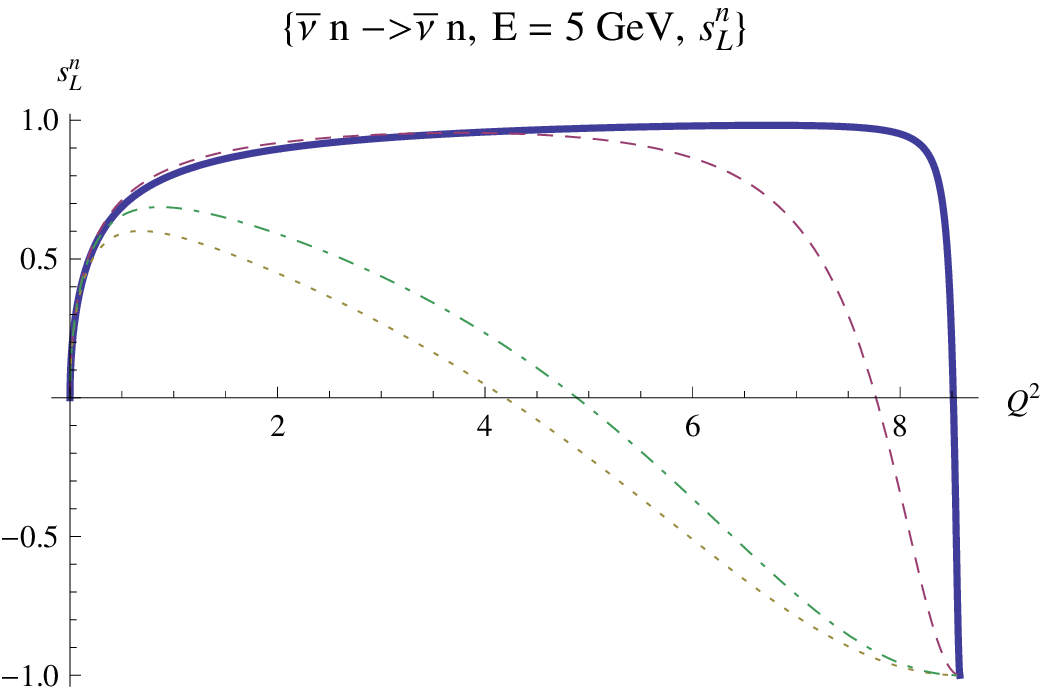,width=75mm}} \caption{The
dependence of  the longitudinal polarization of protons $s_L^p$
(\textit{left}) and  neutrons $s_L^n$ (\textit{right})
 in $\bar\n -N$ scattering on the different choices of $M_A$ and
$g_A^s$,  given in eq. (\ref{parms}). The polarizations are  shown
for
 two neutrino energies: $E=1\, GeV$  (\textit{up}) and  $E=5 \,GeV$
  (\textit{down}).
}\label{sL}
\end{figure}

 The plots on Fig.(\ref{sL})  show the longitudinal polarization.
It is clearly seen that  both for  the  protons and the neutrons it exhibits
a strong sensitivity  to the choice of $M_A$. This sensitivity
  becomes very clearly pronounced  for higher energies, at $Q^2 \geq 1$ $GeV^2$ ($E=5\,GeV$).
For example, at $Q^2=3\, GeV^2$ the proton polarization  changes from
 $s_L^p\simeq 0$ at $M_A=1.0$ to  $s_L^p\simeq 0.7$ at $M_A=1.35$
  with
  almost no sensitivity to $g_A^s$,  the neutron polarization exhibits similar behaviour
  at high energies  (the lower plots in Fig. (\ref{sL})).
   A sensitivity to both the axial and strange axial form factors we find at lower energies
 in  the neutron polarization . For example, at $Q^2\simeq 0.8\, GeV^2$ ($E=1\,GeV$), $s_L^n$ varies from
 -\, 0.4 to +0.4 depending on the choices 1) - 4) (the upper plots in Fig. (\ref{sL})).

\begin{figure}[htb]
   \centerline{ \epsfig{file=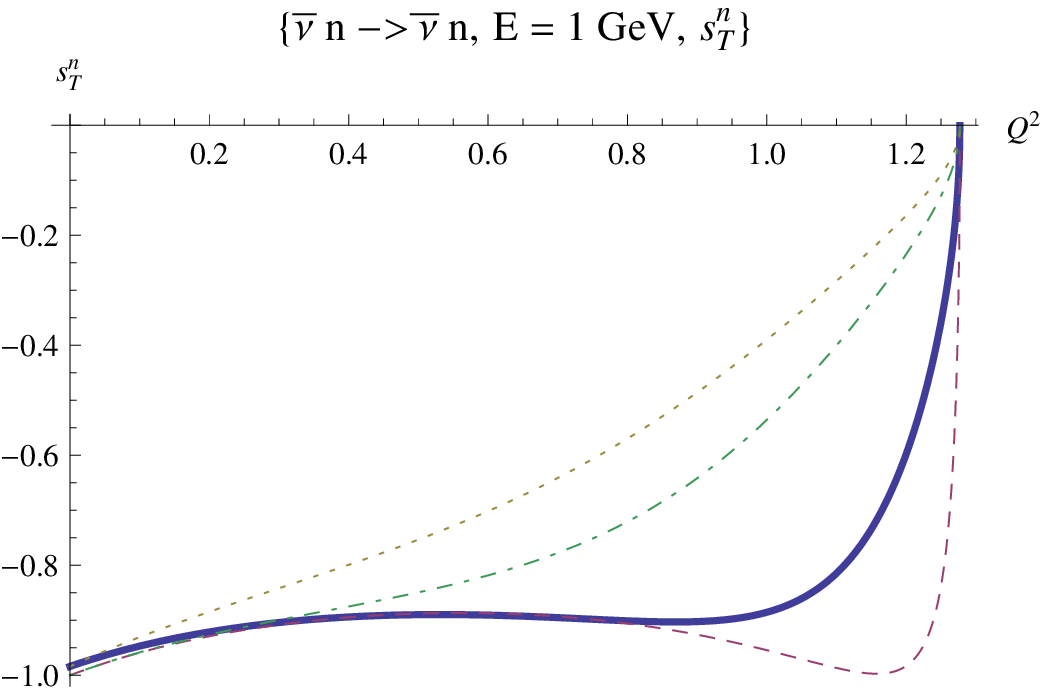,width=75mm}\hspace{0.5cm}\epsfig{file=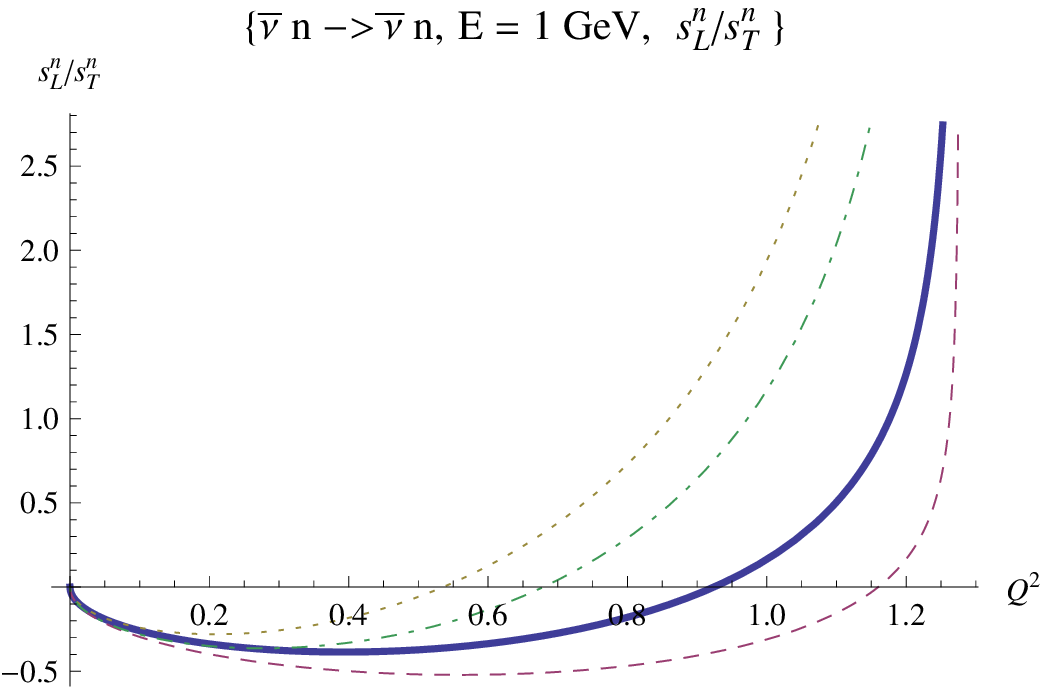,width=75mm}}
    \vspace{10mm}

\centerline{\epsfig{file=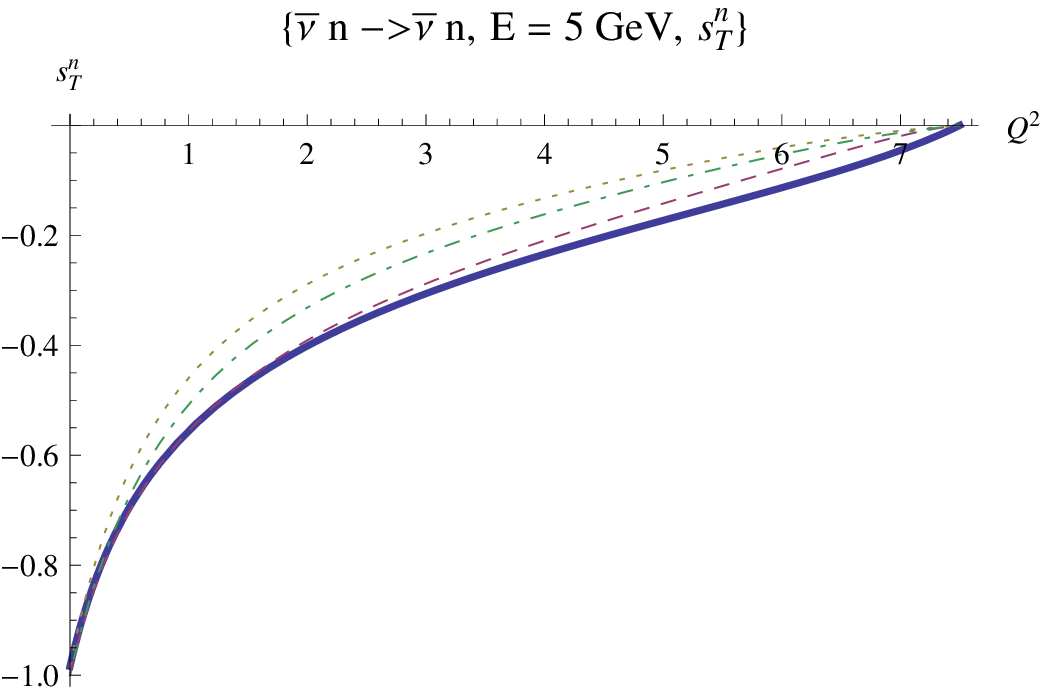,width=75mm}\hspace{0.5cm}\epsfig{file=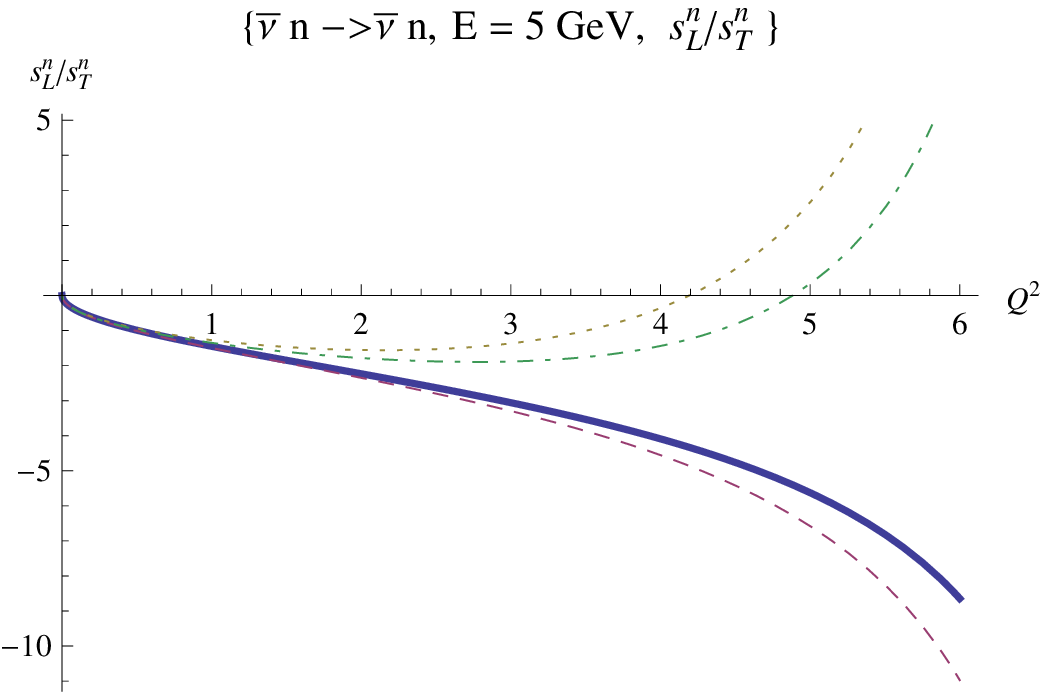,width=75mm}}
 \caption{The dependence of the transverse polarization of the neutron $s_T^n$ (\textit{left})
  and the ratio $s_L^n/s_T^n$ (\textit{right}) in $\bar\n -n$ scattering
 on the different choices of $M_A$ and
$g_A^s$,  given in eq. (\ref{parms}).  The polarizations are  shown for
 two neutrino energies: $E=1\, GeV$  (\textit{up}) and  $E=5 \,GeV$
  (\textit{down}).}\label{nTr}
\end{figure}

On Fig. (\ref{nTr}) we
show the transverse  polarization $s_T^n$ of the neutron and the ratio
$s_L^n/s_T^n$, eq. (\ref{ratio}). At lower
energies $s_T^n$ exhibits a
sensitivity to both the axial and strange axial form factors,
 and almost no sensitivity at $E=5$ GeV (the left plots in Fig. (\ref{nTr})).
 At lower energies the $Q^2$-dependence of $s_L^n/s_T^n$ distinguishes
 all four choices (\ref{parms})
 and becomes sensitive only to $M_A$ at higher energies $Q^2 \geq 3\,GeV^2$.

 As our estimates showed, the transverse polarization of the proton
is very small,  $s_T^p\simeq 0.02 - 0.06$.

\begin{figure}[htb]
  \centerline{\epsfig{file=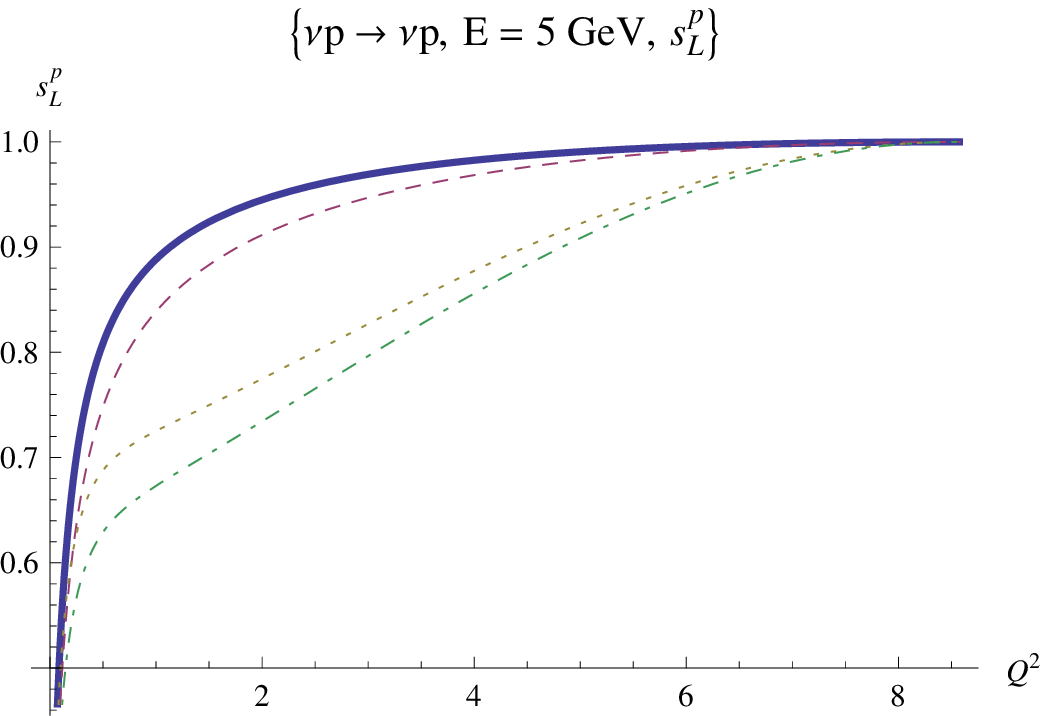,width=75mm}\hspace{0.5cm}\epsfig{file=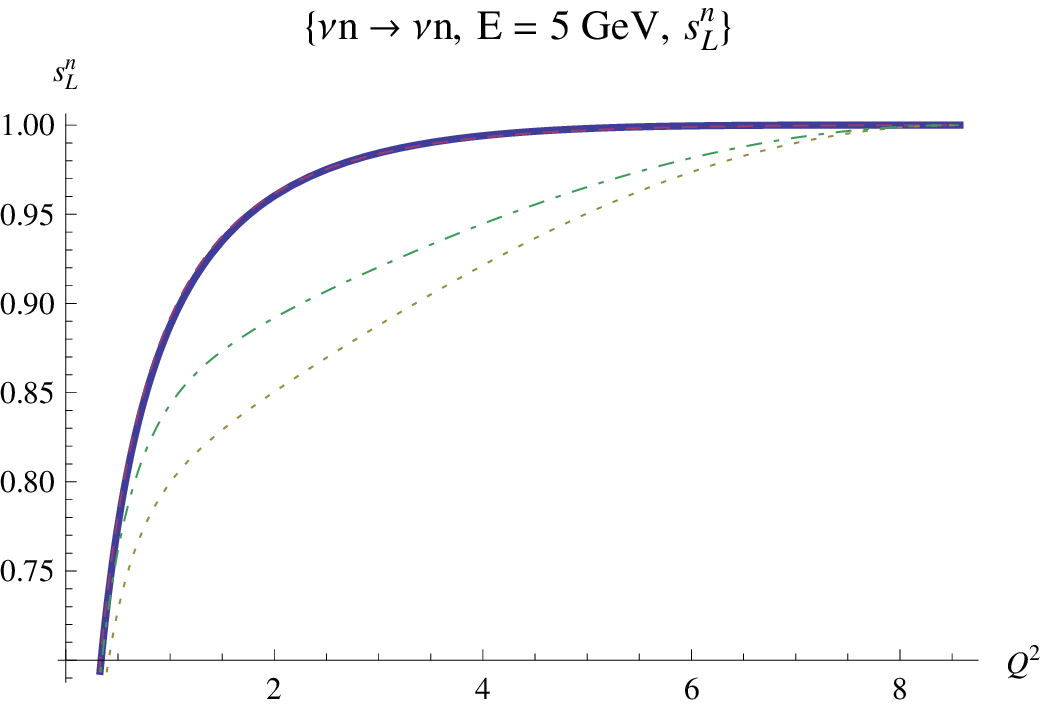,width=75mm}}
  \caption{The longitudinal polarization  in  $\nu -N$ scattering at
 $E=5\,GeV$ of  protons \textit{(left)} and  neutrons \textit{(right)}  for
the different choices of $M_A$ and $g_A^s$ as shown in eq
(\ref{parms}).}\label{Lnu}
\end{figure}

$\bullet$  The polarization in $\n -N$ scattering is big, but shows much weaker sensitivity to the  axial form factors.
For illustration on  Fig. (\ref{Lnu}) we show the longitudinal polarizations at $E= 5\,GeV$,
where the sensitivity is the biggest one.

\begin{figure}[htb]
  \centerline{\epsfig{file=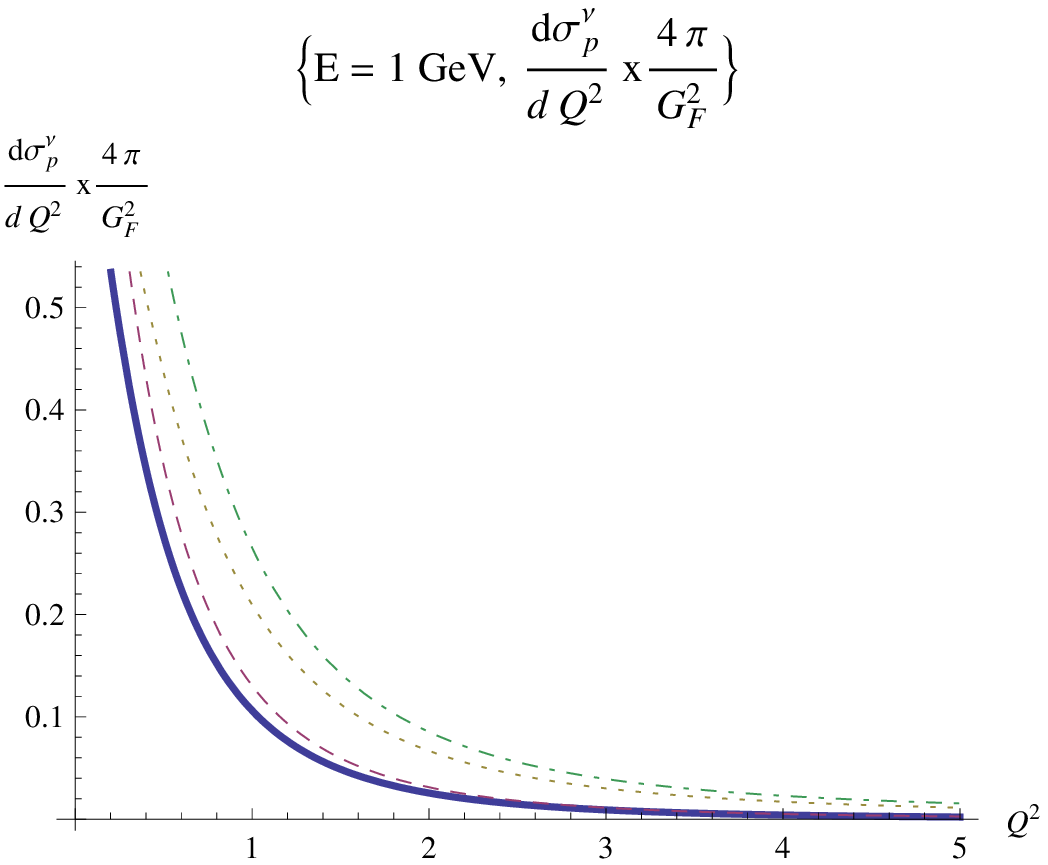,width=75mm}\hspace{0.5cm}\epsfig{file=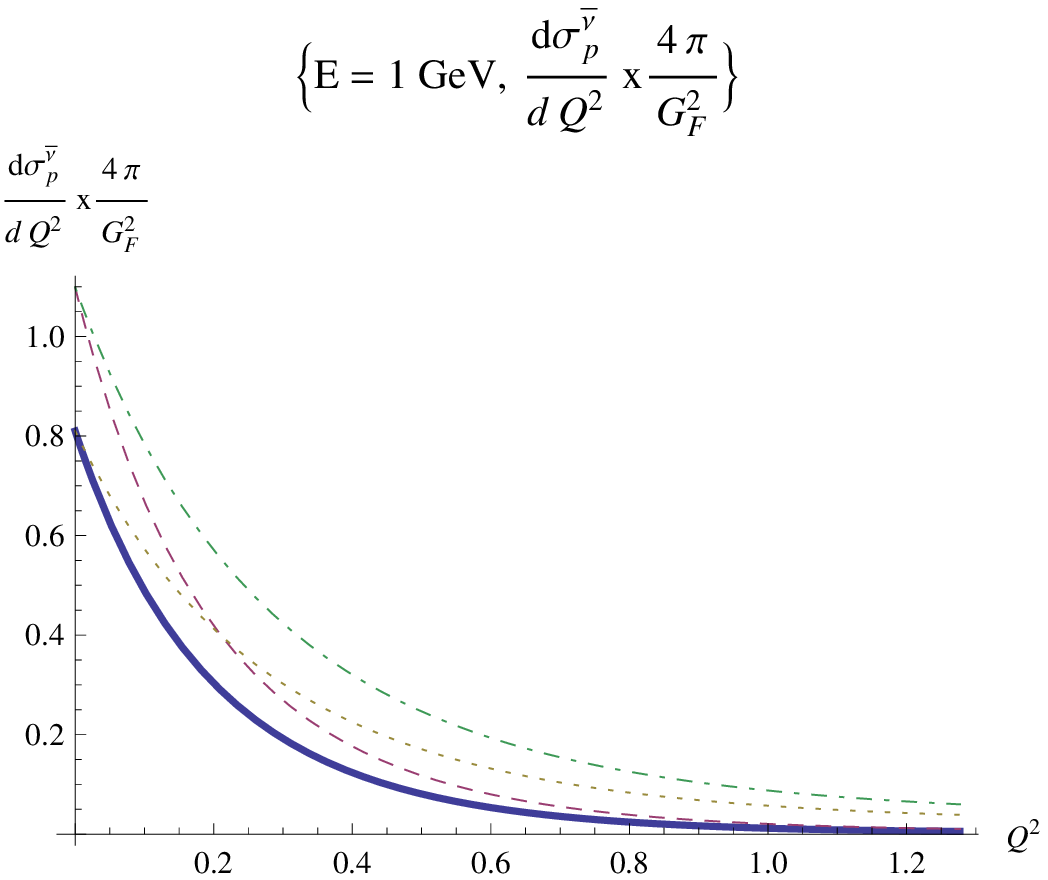,width=75mm}}
  \vspace{10mm}

 \centerline{\epsfig{file=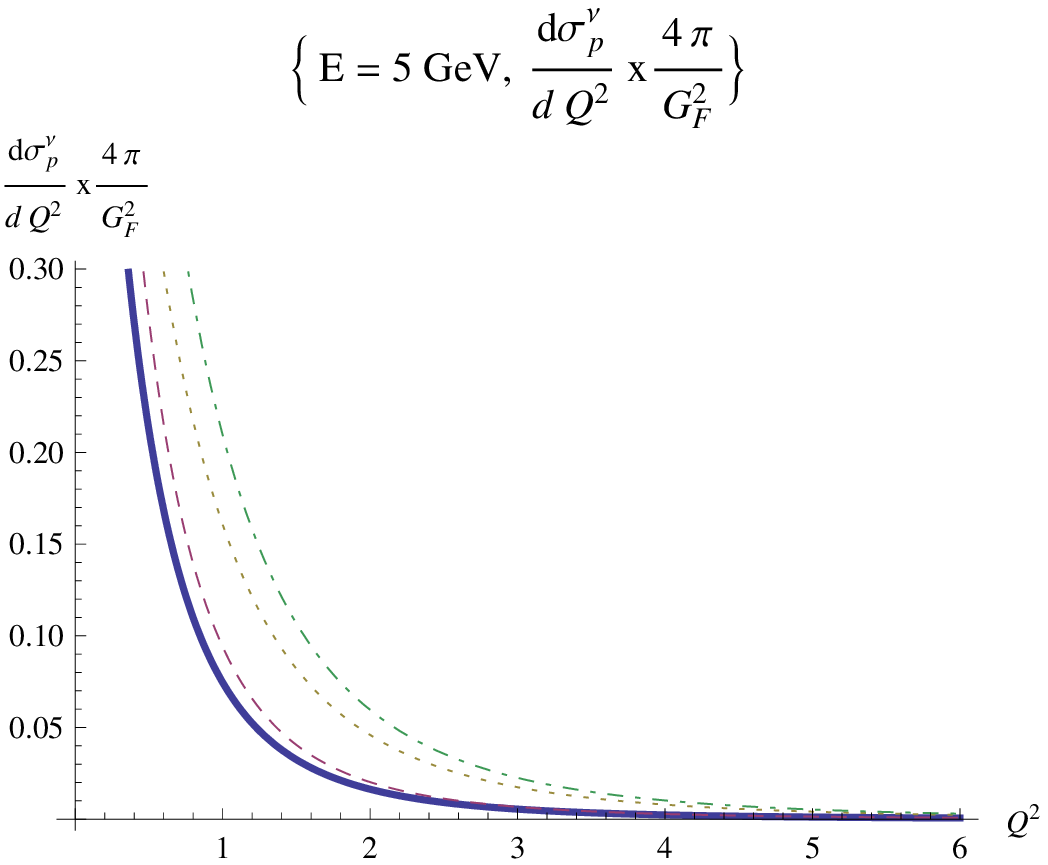,width=75mm}\hspace{0.5cm}\epsfig{file=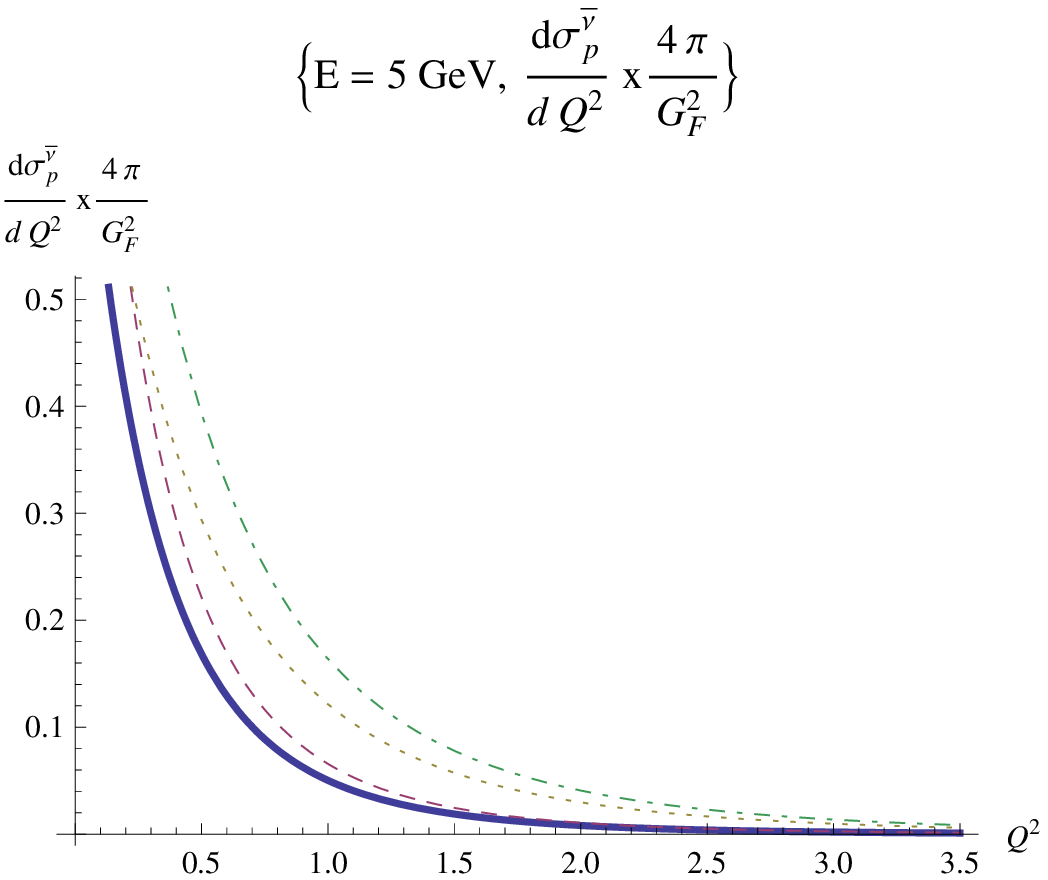,width=75mm}} \caption{The
dependence of $d\sigma_p^{\n}/dQ^2$ (\textit{left}) and $d\sigma_p^{\bar\n}/dQ^2$ (\textit{right})
[\textit{multiplied by $4\pi/G_F^2$}]
on the different choices
of $M_A$ and $g_A^s$, given in eq. (\ref{parms}), at
  $E=5 \,GeV$  for protons
(\textit{up}) and neutrons (\textit{down}).}\label{sigma}
\end{figure}

$\bullet$  On Figs. (\ref{sigma}) we show the differential cross
sections (multiplied by $4\pi/G^{2}_{F}$) for $\nu -p$ and $\bar\nu
-p$ elastic scattering at $E=5 \,GeV^2$ for the same $M_A$ and
$g_A^s$, eq.(\ref{parms}). As compared to the polarization, the
sensitivity to the axial form factors is much weaker.

\section{Conclusions}

In a recent paper \cite{BilChr} we suggested that measurements of the polarization of
the final nucleon in quasi-elastic $\n - N$
scattering could provide
 additional information about the axial form factor $G_A$. Here
 we present the results of the  calculations of the polarization
 of the final nucleons in elastic $\n_\m (\bar\n_\m )- N$
scattering.

The NC form factors, that determine the process, are expressed in terms of the electromagnetic,
axial and  strange vector and axial form factors of the nucleon. We have examined
numerically the sensitivity of the final nucleon polarization to the axial and strange axial form factors.

 Most sensitive to the axial form factors are
the longitudinal polarizations of the proton and  neutron in anti-neutrino nucleon scattering.
This sensitivity increases for higher energies.
Also the value of the polarizations is large.
 In order to measure the longitudinal polarization, like in the case of $e-p$ scattering, a magnetic field must be applied.

The transverse polarization of the proton is extremely small and thus, very difficult to be measured.
This is a consequence of the smallness of the
NC electric form factor of the proton. On the other side the transverse polarization of the neutron is unsuppressed and
 can be  large.  Big and most sensitive to the
axial and strange axial  form factors  is the transverse polarization in $\bar\nu -n$ scattering at small $Q^2\leq 1 \,GeV^2$.
Its determination, however, requires  difficult measurement of left-right asymmetry
 in the scattering of the final neutron.

The measurement of the polarization of the final nucleons in  NC elastic neutrino-nucleon scattering
is a challenge. However,
such measurements could be an important source of information about
axial and strange form factors of  nucleon
in the same way as measurement of the polarization of the final proton  in elastic
$e-p$ scattering  were very important for
the electromagnetic form factors of the proton.
From
our point of view it is worthwhile  to consider a possibility for such measurements short-baseline neutrino experiments.

\section*{Acknowledgments}

S.M. acknowledges support in part by RFBR Grant N 13-02-01442; the
work of E.Ch. is partially supported by a priority Grant between
JINR-Dubna and  Republic Bulgaria, theme 01-3-1070-2009/2013 of the
BLTP.






\begin{thebibliography}{99}


\bibitem{ABMaj} W.M. Alberico, S.M. Bilenky and C. Majoron, Physics Reports {\bf 358}  (2002) 227.

\bibitem{Bodek}A. Bodek {\it et al.} Eur. Phys. J. {\bf C 53}
(2008) 349.
\bibitem{Nomad} V. Lyubushkin {\it et al.} (NOMAD collaboration) Eur. Phys. J. {\bf C 63}
(2009) 355.

\bibitem{Minos} N. Mayer and N. Graf (MINOS-collaboration), AIP Conf. Proc. {\bf 1405}
(2011) 41.

\bibitem{K2K} R. Gran {\it et al.} (K2K collaboration),  Phys. Rev. {\bf D74}
(2006) 052002.

\bibitem{Miniboone} A.A. Aguilar-Arevalo {\it et al.} (MiniBooNE collaboration),Phys. Rev. {\bf D81}
(2010) 092005

\bibitem{Martini1} M. Martini {\it et al.},  Phys. Rev. {\bf C81}
(2010) 045502 (arXiv:1002.4538)

\bibitem{Martini2} M. Martini, M. Ericson, G.Chanfray,
arXiv:1202.4745

\bibitem{T2K} Y. Itow et al, (T2K collaboration) arXiv: hep-ex/0106019

\bibitem{MINERVA} D.D.Stancil et al (MINERVA collaboration), Mod. Phys. Lett. {\bf A27} (2012) 125077 (arXiv:1203.2847)

\bibitem{ArgoNeuT} C. Anderson {\it et al} (ArgoNeuT collaboration),
JINST 7 (2012) P10019 (arXiv:1205.6747).

\bibitem{BilChr}  S.M. Bilenky and E.Christova, J. Phys. G: Nucl. Part. Phys. {\bf 40} (2013) 075004 (arXiv:1303.3710).

\bibitem{Perdrisat}    C.F.Perdrisat, V.Punjabi and M. Vanderhaeghen,
 Prog. Part. Nucl. Phys.{\bf 59} (2007) 694 (arXix:hep-ph/0612014).

\bibitem{Arrington}
J. Arrington, K. de Jager, C. F. Perdrisat
J.Phys.Conf.Ser.{\bf 299} (2011) 012002 (arXiv:1102.2463)


\bibitem{Happex}    Z. Ahmed {\it et al.} (The HAPPEX Collaboration), Phys. Rev. Lett. {\bf 108} (2000) 102001.

\bibitem{BNL}L. A. Ahrens {\it et al.} Phys. Rev. {\bf D35} (1987) 785.

\bibitem{PDG} J. Beringer {\it et al.}(Particle Data Group),
Phys. Rev. {\bf D 86} (2012) 010001.

\bibitem{1} Eden T. et al., Phys. Rev. {\bf C50} (1994) 1749

\bibitem{2} Herberg C. et al., Eur. Phys. J. {\bf A5} (1999) 131; Ostrick M. et al., Phys. Rev. Lett. {\bf 83} (1999) 276


\end{thebibliography}
\end{document}